\documentclass[submission,copyright,creativecommons]{eptcs}
\providecommand{\event}{GandALF 2019} 

\usepackage[utf8]{inputenc}
\usepackage[T1]{fontenc}
\usepackage{mathtools}
\usepackage{tikz}
\usetikzlibrary{automata}
\usepackage{amsmath}
\usepackage{amsthm}
\usepackage{amsfonts}
\usepackage{booktabs}
\usepackage{multirow}
\usepackage{wrapfig}
\usepackage{underscore}           
\usetikzlibrary{calc}
\usepackage{fixltx2e}

\theoremstyle{theorem}
\newtheorem{lemma}{Lemma}

\newtheorem{theorem}{Theorem}

\newtheorem{example}{Example}
\newtheorem{problem}{Problem}

\newcommand{\myquot}[1]{``#1''}

\renewcommand{\epsilon}{\varepsilon}
\newcommand{\bigo}[0]{\mathcal{O}}
\newcommand{\pow}[1]{2^{#1}}
\newcommand{\cceq}{::=}
\newcommand{\card}[1]{|#1|}
\newcommand{\size}[1]{|#1|}
\newcommand{\set}[1]{\{ #1 \}}
\newcommand{\nats}[0]{\mathbb{N}}

\newcommand{\bool}[0]{\mathbb{B}}
\renewcommand{\implies}{\rightarrow}
\newcommand{\cl}{\mathrm{cl}}
\newcommand{\ttrue}{\mathtt{tt}}
\newcommand{\ffalse}{\mathtt{ff}}

\newcommand{\init}[0]{I}

\newcommand{\aut}{\mathfrak{A}}
\newcommand{\autb}{\mathfrak{B}}

\newcommand{\pref}[2]{#1[0,#2)}
\newcommand{\suff}[2]{#1[#2,\infty)}

\newcommand{\ltl}{\text{LTL}}

\newcommand{\prompt}{\text{Prompt-LTL}}
\newcommand{\ldl}{\text{LDL}}
\newcommand{\rltl}{\text{rLTL}(\ensuremath{\Boxdot, \Diamonddot})}

\newcommand{\rprompt}{\text{rPrompt-LTL}}
\newcommand{\rldl}{\text{rLDL}}
\newcommand{\pldl}{\text{PLDL}}
\newcommand{\pltl}{\text{PLTL}}

\newcommand{\rpromptldl}{\text{rPrompt-LDL}}
\newcommand{\promptldl}{\text{Prompt-LDL}}

\newcommand{\prompteval}[0]{V^{\textsc{p}}}
\newcommand{\ldleval}[0]{V^{\textsc{d}}}
\newcommand{\rltleval}[0]{V^{\textsc{r}}}

\newcommand{\rprompteval}[0]{V^{\textsc{rp}}}
\newcommand{\rldleval}[0]{V^{\textsc{rd}}}

\newcommand{\robRexp}{\mathcal{R}^{\textsc{rd}}}

\newcommand{\conc}{\,;}

\newcommand{\tval}{\beta}

\newcommand{\halfthinspace}{{\kern .08333em}}

\newcommand{\ddiamond}[1]{\langle\/ #1 \/\rangle\,}

\newcommand{\bbox}[1]{[\halfthinspace#1\halfthinspace]\,}

\newcommand{\ddiamonddot}[1]{\langle\/ \cdot#1\cdot \/\rangle\,}

\newcommand{\promptddiamonddot}[1]{\langle\/ \cdot#1\cdot \/\rangle\textsubscript{\textbf{p}}\,}

\newcommand{\bboxdot}[1]{[\cdot#1\cdot]\,}

\tikzset{robust/.style={line width=.16ex,line join=round}}

\let\Box\relax
\DeclareMathOperator{\Box}{%
	\text{%
		\tikz[baseline]{%
    			\draw[robust] (0ex,-.1ex) -- (0ex, 1.4ex) -- (1.5ex, 1.4ex) -- (1.5ex, -.1ex) -- cycle;%
		}%
	}%
}

\DeclareMathOperator{\Boxdot}{%
	\text{%
		\tikz[baseline]{%
    			\draw[robust] (0ex, -.1ex) -- (0ex, 1.4ex) -- (1.5ex, 1.4ex) -- (1.5ex, -.1ex) -- cycle;%
	    		\fill (.75ex, .65ex) circle (.15ex);%
    		}%
	}%
}

\let\Diamond\relax
\DeclareMathOperator{\Diamond}{%
	\text{%
		\tikz[baseline]{%
			\draw[robust] (0ex,.6ex) -- (.95ex, 1.55ex) -- (1.9ex, .6ex) -- (.95ex, -.35ex) -- cycle;%
		}%
	}%
}

\DeclareMathOperator{\Diamonddot}{%
	\text{%
		\tikz[baseline]{%
			\draw[robust] (0ex,.6ex) -- (.95ex, 1.55ex) -- (1.9ex, .6ex) -- (.95ex, -.35ex) -- cycle;%
			\fill (.95ex, .6ex) circle (.15ex);%
		}%
	}%
}

\DeclareMathOperator{\Diamondprompt}{%
	\text{%
		\tikz[baseline]{%
			\draw[robust] (0ex,.6ex) -- (.95ex, 1.55ex) -- (1.9ex, .6ex) -- (.95ex, -.35ex) -- cycle;%
		}%
	}%
	\textsubscript{\textbf{p}}%
}

\DeclareMathOperator{\Diamondpromptdot}{%
	\text{%
		\tikz[baseline]{%
			\draw[robust] (0ex,.6ex) -- (.95ex, 1.55ex) -- (1.9ex, .6ex) -- (.95ex, -.35ex) -- cycle;%
			\fill (.95ex, .6ex) circle (.15ex);%
		}%
	}%
	\textsubscript{\textbf{p}}%
}

\newcommand{\itotruthvalue}[1]{0^{#1-1}1^{5-#1}} 

\newcommand{\np}{\textsc{{NP}}}

\newcommand{\pspace}{\textsc{{PSpace}}}

\newcommand{\twoexp}{\textsc{{2ExpTime}}}

\newcommand{\sys}{\mathcal{S}}
\newcommand{\trace}{\lambda}
\newcommand{\game}{\mathcal G}
\newcommand{\ggraph}{G}

\title{Robust, Expressive, and Quantitative Linear Temporal Logics: Pick any Two for Free}

\author{Daniel Neider
\institute{Max Planck Institute for Software Systems\\ 67663 Kaiserslautern, Germany}
\email{neider@mpi-sws.org}
\and
Alexander Weinert
\institute{German Aerospace Center (DLR)\\ Simulation and Software Technology \\ 51147 Cologne, Germany}
\email{alexander.weinert@dlr.de}
\and
Martin Zimmermann
\institute{University of Liverpool\\ Liverpool L69 3BX, United Kingdom}
\email{martin.zimmermann@liverpool.ac.uk}
}

\def\titlerunning{Robust, Expressive, and Quantitative Linear Temporal Logics}
\def\authorrunning{D.\ Neider,  A.\ Weinert \&  M.\ Zimmermann}

\begin{document}
\maketitle

\begin{abstract}
Linear Temporal Logic (LTL) is the standard specification language for reactive systems and is successfully applied in industrial settings.
However, many shortcomings of LTL have been identified in the literature, among them the limited expressiveness, the lack of quantitative features, and the inability to express robustness.
There is work on overcoming these shortcomings, but each of these is typically addressed in isolation.
This is insufficient for applications where all shortcomings manifest themselves simultaneously.

Here, we tackle this issue by introducing logics that address more than one shortcoming.
To this end, we combine the logics Linear Dynamic Logic, Prompt-LTL, and robust LTL, each addressing one aspect, to new logics.
For all combinations of two aspects, the resulting logic has the same desirable algorithmic properties as plain LTL.
In particular, the highly efficient algorithmic backends that have been developed for LTL are also applicable to these new logics. 
Finally, we discuss how to address all three aspects simultaneously.
\end{abstract}

\section{Introduction}
\label{sec-intro}
Linear Temporal Logic ($\ltl$)~\cite{Pnueli77} is amongst the most prominent and most important specification languages for reactive systems, e.g., non-terminating controllers interacting with an antagonistic environment. Verification of such systems against $\ltl$ specifications is routinely applied in industrial settings nowadays~\cite{EisnerFismanPSL,Fix08}. Underlying this success story is the exponential compilation property~\cite{VardiWolper94}: every $\ltl$ formula can be effectively translated into an equivalent Büchi automaton of exponential size (and it turns out that this upper bound is tight). In fact, almost all verification algorithms for $\ltl$ are based on this property, which is in particular true for the popular polynomial space model checking algorithm and the doubly-exponential time synthesis algorithms. Other desirable properties of $\ltl$ include its compact and variable-free syntax and its intuitive semantics.

Despite the success of $\ltl$, a plethora of extensions of $\ltl$ have been studied, all addressing individual and specific shortcomings of $\ltl$, e.g., its limited expressiveness, its lack of quantitative features, and its inability to express robustness. Commonly, extensions of $\ltl$ as described above are only studied in isolation---the logics are either more expressive, or quantitative, or robust.
One notable exception is Parametric $\ldl$ ($\pldl$)~\cite{FaymonvilleZimmermann17}, which adds quantitative operators and increased expressiveness while maintaining the exponential compilation property and intuitive syntax and semantics.
In practical settings, however, it does not suffice to address one shortcoming of $\ltl$ while ignoring the others.
Instead, one needs a logic that combines multiple extensions while still maintaining the desirable properties of $\ltl$.
The overall goal of this paper is, hence, to bridge this gap, thereby enabling expressive, quantitative, and robust verification and synthesis. 

It is a well-known fact that $\ltl$ is strictly weaker than Büchi automata, i.e., it does not harness the full expressive power of the algorithmic backends.
Thus, increasing the expressiveness of $\ltl$ has generated much attention~\cite{LeuckerSanchez07,Vardi11,VardiWolper94,Wolper83} as it can be easily exploited:
as long as the new logic also has the exponential compilation property, the same optimized backends as for $\ltl$ can be used.
A prominent and recent example of such an extension that yields the full expressive power of Büchi automata is Linear Dynamic Logic ($\ldl$)~\cite{Vardi11}, which adds to $\ltl$ temporal operators guarded by regular expressions. 
As an example, consider the specification \myquot{p holds at every even time point, but may or may not hold at odd time points}. It is well-known that this property is not expressible in $\ltl$, as $\ltl$, intuitively, is unable to count modulo a fixed number. However, the specification is easily expressible in $\ldl$ as $\bbox{r} p$, where $r$ is the regular expression~$(\ttrue \cdot \ttrue)^*$. The formula requires $p$ to be satisfied at every position~$j$ such that the prefix up to position~$j$ matches the regular expression~$r$ (which is equivalent to $j$ being even), i.e., $\ttrue$ is an atomic regular expression that matches every letter. In this work, we consider $\ldl$ instead of the alternatives cited above for its conceptual simplicity: $\ldl$ has a simple and variable-free syntax based on regular expressions as well as intuitive semantics (assuming some familiarity with regular expressions).

Another serious shortcoming of $\ltl$ (and $\ldl$) is its inability to adequately express timing bounds. For example, consider the specification \myquot{every request~$q$ is eventually answered by a response~$p$}, which is expressed in $\ltl$ as $\Box (q \rightarrow \Diamond p)$.
It is satisfied, even if the waiting time between requests~$q$ and responses~$p$ diverges to infinity, although such a behavior is typically undesired. Again, a long line of research has addressed this second shortcoming of $\ltl$~\cite{AlurEtessamiLaTorrePeled01,FaymonvilleZimmermann17,Koymans90,KupfermanPitermanVardi09,Zimmermann18}.
The most basic one is $\prompt$~\cite{KupfermanPitermanVardi09}, which adds the prompt-eventually operator~$\Diamondprompt$ to $\ltl$.
The semantics is now defined with an additional parameter~$k$, which bounds the scope of $\Diamondprompt$: $\Box (q \rightarrow \Diamondprompt p)$ requires every request~$q$ to be answered within $k$ steps, when evaluated with respect to $k$.
The resulting logic is a quantitative one: either one quantifies the parameter~$k$ existentially and obtains a boundedness problem, e.g., \myquot{is there a bound~$k$ such that every request can be answered within $k$ steps}, or one even aims to determine the optimal bound~$k$.
Again, $\prompt$ retains the desirable properties of $\ltl$, i.e., the exponential compilation property as well as intuitive syntax and semantics. Furthermore, $\prompt$ captures the technical core of the alternatives cited above, e.g., decision problems for the more general logic~$\pltl$~\cite{AlurEtessamiLaTorrePeled01} can be reduced to those for $\prompt$. For these reasons, we study $\prompt$ in this work.

Finally, a third line of extensions of $\ltl$ is concerned with the concept of robustness, which is much harder to formalize. This is reflected by a multitude of incomparable notions of robustness in verification~\cite{DBLP:conf/cav/BloemCGHJ10,DallalNeiderTabuada16,DBLP:conf/formats/DonzeM10,DBLP:conf/acsd/DoyenHLN10,DBLP:journals/tcs/FainekosP09,DBLP:conf/rtss/MajumdarS09,NeiderWeinertZimmermann18,DBLP:journals/tac/TabuadaCRM14,TabuadaNeider16}.
Here, we are interested in robust $\ltl$ ($\text{r}\ltl$)~\cite{TabuadaNeider16}, which equips $\ltl$ with a five-valued semantics that captures different degrees of violations of universal specifications.
As an example, consider the specification \myquot{if property $\varphi$ always holds true, then property $\psi$ also always holds true}, which is expressed in $\ltl$ as $\Box \varphi \rightarrow \Box \psi$ and is typical for systems that have to interact with an antagonistic environment.
In classical semantics, the whole formula is satisfied as soon as the assumption~$\varphi$ is violated once, even if the guarantee~$\psi$ is violated as well.
By contrast, the semantics of robust $\ltl$ ensures that the degree of the violation of $\Box \psi$ is always proportional to the degree of the violation of $\Box \varphi$.
To this end, the degree of a violation of a property $\Box \varphi$ is expressed by five different truth values: either $\varphi$ always holds, or $\varphi$ is violated only finitely often, violated infinitely often, violated almost always, or violated always.
Again, robust $\ltl$ has the exponential compilation property and an intuitive syntax (though its semantics is more intricate). In this work, we consider robust $\ltl$, as it is the first logic that intrinsically captures the notion of robustness in $\ltl$. In particular, formulas of robust $\ltl$ are evaluated over traces with Boolean truth values for atomic propositions and do not require non-Boolean assignments, which are often hard to determine in real-life applications.

	
	\subsubsection*{Our Contributions}
We develop logics that address more than one shortcoming of $\ltl$ at a time. See Figure~\ref{fig:logics} for an overview.

\begin{wrapfigure}{R}{.45\textwidth} \centering
\vspace{-.4cm}
	\begin{tikzpicture}[thick]
		
		\draw[rounded corners,fill=black!10,draw=white]
			(-3.3, .5) |- (.9,2.25) |- (3.1,3.5) |- (0,.5) -- cycle;
	
		\node[align=center] (ltl) at (0,0.8) {\ltl};
		
		\begin{scope}[shift={(0,1.75)}]
			\node[align=center] (rltl) at (-2.25,0) {\rltl};
			\node[align=center] (promptltl) at (0,0) {\prompt};
			\node[align=center] (ldl) at (2,0) {\ldl};
		\end{scope}
		
		\begin{scope}[shift={(0,3)}]
			\node[align=center] (rpromptltl) at (-2.1,0) {\rprompt};
			\node[align=center] (rldl) at (0,0) {\rldl};
			\node[align=center] (promptldl) at (2,0) {\promptldl};
		\end{scope}
		
		\node[align=center] (rpromptldl) at (0,4.2) {\rpromptldl};
		
		\path[-stealth,]
			(ltl)
				edge[dashed] (rltl)
				edge[dashed] (promptltl)
				edge[dashed] (ldl)
			(rltl)
				edge[out=90,in=-90] (rpromptltl)
				edge[out=30,in=-90] (rldl)
			(promptltl)
				edge[out=150,in=-90] (rpromptltl)
				edge[dashed,out=30,in=-90] (promptldl)
			(ldl)
				edge[out=135,in=-90] (rldl)
				edge[dashed] (promptldl)
			(rpromptltl)
				edge[out=30,in=-90] (rpromptldl)
			(rldl)
				edge (rpromptldl)
			(promptldl)
				edge[out=150,in=-90] (rpromptldl);
	\end{tikzpicture}
	\caption{The logics studied in this work. Existing logics and influences are marked gray with dashed arrows.}
	\label{fig:logics}
\end{wrapfigure}

	In Section~\ref{sec-rprompt}, we ``robustify'' $\prompt$. More precisely, we introduce a novel logic, named $\rprompt$, by extending the five-valued semantics from robust $\ltl$ to $\prompt$. Our main result here shows that $\rprompt$ retains the exponential compilation property. 
Then, in Section~\ref{sec-rldl}, we ``robustify'' $\ldl$: we introduce a novel logic, named $\rldl$, by lifting the five-valued semantics of robust $\ltl$ to $\ldl$. Our main result shows that $\rldl$ also retains the exponential compilation property. 
Hence, one can indeed combine any two of the three extensions of $\ltl$ while still preserving the desirable algorithmic properties of $\ltl$. In particular, let us stress again that all highly sophisticated algorithmic backends developed for $\ltl$ are applicable to these novel logics as well, e.g., we show that the verification problem and the synthesis problem for each of these logics is solvable without an (asymptotic) increase in complexity.

Tabuada and Neider gave two proofs showing that robust $\ltl$ has the exponential compilation property. The first one presented a translation of robust $\ltl$ into equivalent Büchi automata of exponential size while the second one is based on a polynomial translation of robust $\ltl$ into (standard) $\ltl$, which is known to be translatable into equivalent Büchi automata of exponential size. We refer to those two approaches as the \emph{direct} approach and the \emph{reduction-based} approach. To obtain our results mentioned above, we need to generalize both. 
To prove the exponential compilation property for $\rldl$, we generalize the direct approach by exhibiting a direct translation of $\rldl$ into Büchi automata via alternating automata. In contrast, to prove the exponential compilation property for $\rprompt$, we present a generalization of the reduction-based approach translating $\rprompt$ into equivalent $\prompt$ formulas of linear size, which have the exponential compilation property.  
	
	 Finally, in Section~\ref{sec-towardsrpldl}, we discuss the combination of all three aspects. Recall that we present a direct translation to automata for $\rldl$ and a reduction-based one for $\rprompt$. For reasons we discuss in Section~\ref{sec-towardsrpldl}, it is challenging to develop a reduction from $\rldl$ to $\ldl$ or a direct translation for $\rprompt$ that witness the exponential compilation property. Hence, both approaches seem inadequate to deal with the combination of all three extensions. Ultimately, we leave the question of whether the logic combining all three aspects has the exponential compilation property for future work.

Proofs omitted due to space restrictions can be found in the full version~\cite{fullversion}.

\section{Preliminaries}
\label{sec-prel}
We denote the non-negative integers by~$\nats$, the set~$\set{0,1}$ of Boolean truth values by~$\bool$, and the power set of~$S$ by~$\pow{S}$.
By convention, we have $\min \emptyset = 1$ and $\max \emptyset = 0$ when the operators range over subsets of $\bool$. Following Tabuada and Neider~\cite{TabuadaNeider16}, the set of truth values for robust semantics is $\bool_4 = \set{0000, 0001, 0011, 0111, 1111}$, which are ordered by $0000 \prec 0001 \prec 0011 \prec 0111 \prec 1111$.
We write $\preceq$ for the non-strict variant of $\prec$ and define $\min \emptyset = 1111$ and $\max \emptyset = 0000$ when the operators range over subsets of $\bool_4$.

Throughout this work, we fix a finite non-empty set~$P$ of atomic propositions.
 For a set $A \subseteq P$ and a propositional formula~$\phi$ over $P$, we write $A \models \phi$ if the variable valuation mapping elements in $A$ to $1$ and elements in $P\setminus A$ to $0$ satisfies $\phi$.
A trace (over $P$) is an infinite sequence~$w \in (\pow{P})^\omega$. Given a trace~$w = w(0) w(1) w(2) \cdots$ and a position~$j \in \nats$, we define $\pref{w}{j} = w(0) \cdots w(j-1)$ and $\suff{w}{j} = w(j) w(j+1) w(j+2) \cdots$, i.e., $\pref{w}{j}$ is the prefix of length~$j$ of $w$ and $\suff{w}{j}$ the remaining suffix. In particular, $\pref{w}{0}$ is empty  and $\suff{w}{0}$ is $w$.
 
 Our work is based on three logics, Robust Linear Temporal Logic ($\rltl$)~\cite{TabuadaNeider16}, Linear Dynamic Logic ($\ldl$)~\cite{Vardi11}, and Prompt Linear Temporal Logic ($\prompt$)~\cite{KupfermanPitermanVardi09}, which we briefly review in the following three subsections. More formal definitions can be found in in the original publications introducing these logics and in the full version of this work~\cite{fullversion}.
 
 We define the semantics of all these logics by evaluation functions~$V$ mapping a trace, a formula, and a bound (in the case of a quantitative logic) to a truth value. This is prudent for robust semantics, hence we also use this approach for the other logics, which are typically defined via satisfaction relations. In particular, $\rltleval$, $\ldleval$, and $\prompteval$ denote the evaluation functions of $\rltl$, $\ldl$, and $\prompt$, respectively. Nevertheless, our definitions here are equivalent to the original definitions.



\subsection{Robust Linear Temporal Logic}
\label{subsec-briefrltl}
The main impetus behind the introduction of robust $\ltl$ was the need to capture the concept of robustness in temporal logics. As a first motivating example consider the $\ltl$ formula~$\Box p$, stating that $p$ holds at every position. Consequently, the formula is violated if there is a single position where $p$ does not hold. However, this is a very \myquot{mild} violation of the property and there are much more \myquot{severe} violations. As exhibited by Tabuada and Neider, there are four canonical \emph{degrees} of violation of $\Box p$: (i) $p$ is violated at finitely many positions, (ii) $p$ is violated at infinitely many positions, (iii) $p$ is violated at all but finitely many positions, and (iv) $p$ is violated at all positions. These first three degrees are captured by the $\ltl$ formulas $\Diamond\Box p$, $\Box \Diamond p$, and $\Diamond p$, which are all weakenings of $\Box p$. All five possibilities, satisfaction and four degrees of violation, are captured in robust $\ltl$ by the truth values 
\[ 1111 \succ 0111 \succ 0011 \succ 0001 \succ 0000 \]
 introduced above. By design, the formula~$\Boxdot p$ of robust $\ltl$\footnote{Following the precedent for robust $\ltl$, we use dots to distinguish operators of robust logics from those of classical logics troughout the paper.} has
\begin{itemize}
	\item truth value $1111$ on all traces where $p$ holds at all positions,
	\item truth value $0111$ on all traces where $p$ holds at all but finitely many positions,
	\item truth value $0011$ on all traces where $p$ holds at infinitely many positions and does not hold at infinitely many positions,
	\item truth value $0001$ on all traces where $p$ only holds at finitely many positions, and 
	\item truth value $0000$ on all traces where $p$ holds at no position.
\end{itemize}

As a further example, consider the formula~$\Boxdot p \rightarrow \Boxdot q$. For this formula, the robust semantics captures the intuition described in the introduction: the implication is satisfied (i.e., has truth value $1111$), if  the degree of violation of the property~\myquot{always q} is at most the degree of violation  of the property~\myquot{always p}. Thus, if $p$ is violated finitely often, then $q$ may also be violated finitely often (but not infinitely often) while still satisfying the implication.

Conjunction and disjunction are defined as usual using minimization and maximization relying on the order indicated above while negation is based on the intuition that $1111$ represents satisfaction and all other truth values represent degrees of violation. Hence, a negation~$\neg \varphi$ is satisfied (i.e., has truth value $1111$), if $\varphi$ has truth value less than $1111$, and it is violated (i.e., has truth value $0000$) if $\varphi$ has truth value~$1111$. Finally, the semantics of the eventually operator is defined as usual, i.e., the truth value of $\Diamonddot \varphi$ on $w$ is the maximal truth value that is assumed by $\varphi$ on some suffix of $w$. 

This intuition is formalized in the evaluation function~$\rltleval$, which maps a trace~$w \in (\pow{P})^\omega$ and an $\rltl$ formula $\varphi$ to a truth value~$\rltleval(w,\varphi)$ in $\bool_4 $~\cite{TabuadaNeider16}. Note that, for the sake of simplicity, we restrict ourselves to $\rltl$, i.e., robust $\ltl$ with the always and eventually operators, but without next, until, and release. We comment on the effect of this restriction when defining the combinations of logics.

\subsection{Linear Dynamic Logic}
\label{subsec-briefldl}
The logic $\ldl$ has only two temporal operators, $\ddiamond{r}$ and $\bbox{r}$, which can be understood as guarded variants of the classical eventually and always operators from $\ltl$, respectively. Both are guarded by regular expressions~$r$ over the atomic propositions that may contain tests, which are again $\ldl$ formulas. These two operators together with Boolean connectives capture the full expressive power of the $\omega$-regular expressions, i.e., $\ldl$ exceeds the expressiveness of $\ltl$. 

Formally, a formula~$\ddiamond{r}\varphi$ is satisfied by a trace~$w$, if there is some~$j$ such that the prefix~$\pref{w}{j}$ matches the regular expression~$r$ and the corresponding suffix~$\suff{w}{j}$ satisfies $\varphi$. Dually, a formula~$\bbox{r}\varphi$ is satisfied by a trace~$w$ if for every~$j$ with $\pref{w}{j}$ matching $r$, $\suff{w}{j}$ satisfies $\varphi$. Thus, while the classical eventually and always operator range over all positions, the operators of $\ldl$ range only over those positions whose induced prefix matches the guard of the operator.

The semantics of $\ldl$ is captured by the evaluation function~$\ldleval$ mapping a trace~$w \in (\pow{P})^\omega$ and an $\ldl$ formula $\varphi$ to a truth value~$\ldleval(w,\varphi)$ in $\bool$~\cite{GiacomoVardi13,Vardi11}.

\subsection{Prompt Linear Temporal Logic}
\label{subsec-briefprompt}
To express timing constraints, the logic $\prompt$ adds the prompt-eventually operator~$\Diamondprompt$ to $\ltl$. For technical reasons~\cite{AlurEtessamiLaTorrePeled01}, this requires to disallow negation and implication. Intuitively, the new operator requires its argument to be satisfied within a bounded number of steps. 

Thus, the semantics is given by an evaluation function~$\prompteval$ that maps a trace~$w \in (\pow{P})^\omega$, a $\prompt$ formula $\varphi$, and a bound~$k \in\nats$ to a truth value~$\rltleval(w,k,\varphi)$ in $\bool$~\cite{KupfermanPitermanVardi09}.  This function is defined as usual for all Boolean and standard temporal operators (ignoring the bound~$k$), while a formula~$\Diamondprompt \varphi$ is satisfied with respect to the bound~$k$ if $\varphi$ holds within the next $k$ steps, i.e., the prompt-eventually behaves like the classical eventually with a bounded scope.

\section{Robust and Prompt Linear Temporal Logic}
\label{sec-rprompt}
We begin our treatment of combinations of the three basic logics by introducing robust semantics for $\prompt$, obtaining the logic~$\rprompt$. To this end, we add the prompt-eventually operator to $\rltl$ while disallowing implications and restricting negation to retain decidability (cf.~\cite{AlurEtessamiLaTorrePeled01}). The formulas of $\rprompt$ are given by
\[\varphi \cceq p \mid \neg p \mid \varphi \wedge \varphi \mid \varphi \vee \varphi \mid \Diamonddot \varphi 
  \mid \Boxdot \varphi\mid \Diamondpromptdot \varphi,
\]%
where $p$ ranges over the set $P$ of atomic propositions. 
The size~$\size{\varphi}$ of a formula~$\varphi$ is the number of its distinct subformulas. 

The semantics of $\rprompt$ is given by an evaluation function~$\rprompteval$ mapping a trace~$w$, a bound~$k$ for the prompt-eventuallies, and a formula~$\varphi$ to a truth value in $\bool_4$.  To simplify our notation, we write $\rprompteval_i(w,k,\varphi)$ for $i \in \set{1,2,3,4}$ to denote the $i$-th bit of $\rprompteval(w,k,\varphi)$, i.e., 
\[
\rprompteval(w,k,\varphi) = \rprompteval_1(w,k,\varphi)\rprompteval_2(w,k,\varphi)\rprompteval_3(w,k,\varphi)\rprompteval_4(w,k,\varphi).
\]
The semantics of Boolean connectives as well as of the eventually and always operators is defined as for robust $\ltl$. The motivation behind these definitions is carefully and convincingly discussed by Tabuada and Neider~\cite{TabuadaNeider16}. The semantics of the prompt-eventually operator bounds its scope to the next $k$ positions as in classical $\prompt$~\cite{KupfermanPitermanVardi09}.
\begin{itemize}
\item $\rldleval(w,k,p) =\begin{cases}
	1111 &\text{if $p \in w(0)$,}\\
	0000 &\text{if $p \notin w(0)$},
\end{cases}$ \quad\quad $\bullet \,\,\, \rldleval(w,k,\neg p) =\begin{cases}
	1111 &\text{if $p \notin w(0)$,}\\
	0000 &\text{if $p \in w(0)$},
 \end{cases}$

\item
$\rldleval(w, k,\varphi_0 \wedge \varphi_1) = \min\set{\rldleval(w,k, \varphi_0), \rldleval(w,k,\varphi_1) }$, 
\item $\rldleval(w, k,\varphi_0 \vee \varphi_1) = \max\set{\rldleval(w, k,\varphi_0), \rldleval(w,k,\varphi_1) }$, 
	\item $\rprompteval(w,k,\Diamonddot \varphi) = b_1 b_2 b_3 b_4$ where $b_i = \max_{j \in \nats}\rprompteval_i(\suff{w}{j},k,\varphi )$,\footnote{This definition is equivalent to $\rprompteval(w,k,\Diamonddot \varphi) = \max_{j \in \nats}\rprompteval(\suff{w}{j},k,\varphi )$ due to monotonicity of the truth values, which is closer to the classical semantics of the eventually operator. A similar equivalence holds for $\Diamondprompt\varphi$.} and
	\item $\rprompteval(w,k, \Boxdot\varphi) = b_1 b_2 b_3 b_4$ where
\begin{itemize}
	
	\item $b_1 = \min_{j \in \nats} \rprompteval_1(\suff{w}{j},k,\varphi )$, i.e., $b_1 = 1$ iff $\varphi$ holds always,
	 
	\item $b_2 = \max_{j' \in \nats} \min_{j' \le j} \rprompteval_2(\suff{w}{j},k,\varphi )$, i.e., $b_2 = 1$ iff $\varphi$ holds almost always, 
		
	\item $b_3 = \min_{j' \in \nats} \max_{j' \le j} \rprompteval_3(\suff{w}{j},k,\varphi )$, i.e., $b_3 = 1$ iff $\varphi$ holds infinitely often, and 
	
	\item $b_4 = \max_{j \in \nats} \rprompteval_4(\suff{w}{j},k,\varphi )$ i.e., $b_4 = 1$ iff $\varphi$ holds at least once.

\end{itemize}
\item $\rprompteval(w,k, \Diamondpromptdot \varphi) = b_1 b_2 b_3 b_4$ where $b_i = \max_{0 \le j \le k}\rprompteval_i(\suff{w}{j},k,\varphi )$.
\end{itemize}
It is easy to verify that $\rprompteval(w,k,\varphi)$ is well-defined, i.e., $\rprompteval(w,k,\varphi) \in \bool_4$ for all $w$, $k$, and $\varphi$. 

\begin{example}
\label{example-rprompt}
Consider the formula~$\varphi = \Boxdot \Diamondpromptdot s$, where we interpret occurrences of the atomic proposition~$s$ as synchronizations. Then, the different degrees of satisfaction of the formula express the following possibilities, when evaluating it with respect to $k \in \nats$: (i) the distance between synchronizations is bounded by $k$, (ii) from some point onwards, the distance between synchronizations is bounded by $k$, (iii) there are infinitely many synchronizations, and (iv) there is at least one synchronization. Note that the last two possibilities are independent of $k$, which is explained by simple logical equivalences, e.g., the third possibility reads actually as follows: there are infinitely many positions such that a synchronization occurs within distance~$k$. However, it is easy to see that is equivalent to the property stated above. 
\end{example}

In the next two sections, we solve the model checking problem and the synthesis problem for $\rprompt$. To this end, we translate every $\rprompt$ formula into a sequence of five $\prompt$ formulas that capture the five degrees of satisfaction and violation by making the semantics of the robust always operator explicit. This is a straightforward generalization of the, in the terms of the introduction, reduction-based approach to robust $\ltl$~\cite{TabuadaNeider16}. 

\begin{lemma}
\label{lemma-prltl2pltl}
For every $\rprompt$ formula~$\varphi$ and every $\tval \in \bool_{4}$, there is a $\prompt$ formula~$\varphi_\tval$ of size~$\bigo(\size{\varphi})$ such that $\rprompteval(w, k, \varphi) \succeq \tval$ if and only if $\prompteval(w, k, \varphi_\tval) = 1$.
\end{lemma}

Note that the logic $\rltl$ is not a fragment of $\rprompt$ as we have to disallow negation and implication to retain decidability~\cite{AlurEtessamiLaTorrePeled01}.
Conversely, $\prompt$ is also not a fragment of $\rprompt$ as we omitted the next, until, and release operator. However, we present a reduction-based approach from $\rprompt$ to $\prompt$. Thus, one could easily add the additional temporal operators to $\rprompt$ while maintaining the result of Lemma~\ref{lemma-prltl2pltl}. We prefer not to do so for the sake of accessibility and brevity.

\subsection{Model Checking}
\label{subsec-rpromptresults-mc}

Let us now consider the $\rprompt$ model checking problem, which asks whether all executions of a given finite transition system satisfy a given specification expressed as an $\rprompt$ formula with truth value at least $\tval \in \bool_4$.
More formally, we assume the system under consideration to be modeled as a (labeled and initialized) transition system~$\sys = (S, s_\init, E, \lambda)$ over~$P$ consisting of a finite set~$S$ of states containing the initial state~$s_\init$, a directed edge relation~$E \subseteq S \times S$, and a state labeling~$\lambda \colon S \rightarrow 2^P$ that maps each state to the set of atomic propositions that hold true in this state. A path through $\sys$ is a sequence~$\rho = s_0 s_1 s_2 \cdots $ satisfying $s_0 = s_\init$ and $(s_j, s_{j+1}) \in E$ for every $j \in \nats$, and $\Pi_\sys$ denotes the set of all paths through $\sys$. Finally, the trace of a path~$\rho = s_0 s_1 s_2 \cdots \in \Pi_\sys$ is the sequence $\trace(\rho) = \lambda(s_0) \lambda(s_1) \lambda(s_2)\cdots$ of labels induced by $\rho$.
\begin{problem} \label{prob:prLTL-model-checking}
Let $\varphi$ be an $\rprompt$ formula, $\sys$ a transition system, and $\tval \in \bool_4$. Is there a $k \in\nats$ such that $\rprompteval(\trace(\rho), k, \varphi) \succeq \tval$ holds true for all paths~$\rho \in \Pi_\sys$?
\end{problem}

Our solution relies on Lemma~\ref{lemma-prltl2pltl} and on $\prompt$ model checking being in $\pspace$~\cite{KupfermanPitermanVardi09}.

\begin{theorem}
\label{thm-rpltlmodelchecking}
$\rprompt$ model checking is in $\pspace$.
\end{theorem}

We do not claim $\pspace$-hardness because model checking the fragment of $\ltl$ with disjunction, conjunction, always, and eventually operators only (and classical semantics) is $\np$-complete~\cite{AlurLatorre04}. Since this fragment can be embedded into $\rprompt$ (via a translation of this $\ltl$ fragment into $\rprompt$ using techniques similar to those presented by Tabuada and Neider~\cite{TabuadaNeider16} for translating $\ltl$ into $\rltl$), we obtain at least $\np$-hardness for Problem~\ref{prob:prLTL-model-checking}. As we have no next, until, and release operators (by our own volition), we cannot easily claim $\pspace$-hardness. In contrast, the solution of the $\prompt$ model checking problem consists of a reduction to $\ltl$ model checking that introduces until operators (see~\cite{KupfermanPitermanVardi09}). Hence, we leave the fragment mentioned above, for which $\np$ membership is known. However, adding next, until, and release to $\rprompt$ yields a $\pspace$-hard model checking problem. 

\subsection{Synthesis}
\label{subsec-rpromptresults-synt}

Next, we consider the problem of synthesizing reactive controllers from $\rprompt$ specifications. In this context, we rely on the classical reduction from reactive synthesis to infinite-duration two-player games over finite graphs. In particular, we show how to construct a finite-state winning strategy for games with $\rprompt$ winning conditions, which immediately correspond to implementations of reactive controllers.
Throughout this section, we assume familiarity with games over finite graphs~(see, e.g., \cite[Chapter~2]{GraedelThomasWilke02}).

We consider $\rprompt$ games over~$P$, which are triples $\game = (\ggraph, \varphi, \tval)$ consisting of
	 a labeled game graph $\ggraph$,
an $\rprompt$ formula $\varphi$, and a truth value $\tval \in \bool_4$.
A labeled game graph~$\ggraph = (V_0, V_1, E, \lambda)$ consists of a  directed graph $(V_0 \cup V_1, E)$, two finite, disjoint sets of vertices~$V_0$ and~$V_1$, and a function $\lambda \colon V_0 \cup V_1 \to 2^P$ mapping each vertex~$v$ to the set~$\lambda(v)$ of atomic propositions that hold true in $v$.
We denote the set of all vertices by $V = V_0 \cup V_1$ and assume that game graphs do not have terminal vertices, i.e., $\{ v \} \times V \cap E \neq \emptyset$ for each $v \in V$.

As in the classical setting, $\rprompt$ games are played by two players, Player~0 and Player~1, who move a token along the edges of the game graph ad infinitum (if the token is currently placed on a vertex $v \in V_i$, $i \in \{ 0, 1\}$, then Player~$i$ decides the next move). 
The resulting infinite sequence $\rho = v_0 v_1 v_2\cdots \in V^\omega$ of vertices is called a {play} and induces a trace $\lambda(\rho) = \lambda(v_0) \lambda(v_1)\lambda(v_2) \cdots \in (2^P)^\omega$.

A strategy of Player~$0$ is a mapping $f \colon V^\ast V_0 \to V$ that prescribes where to move the token depending on the finite play prefix constructed so far. A play~$v_0v_1v_2 \cdots$ is played according to $f$ if $v_{j+1} = f(v_0 \cdots v_j)$ for every $j$ with $v_j \in V_0$. A strategy $f$ of Player~$0$ is winning from a vertex~$v \in V$ if there is a $k \in \nats$ such that all plays~$\rho$ that start in $v$ and that are played according to $f$ satisfy $\rprompteval(\lambda(\rho), k, \varphi) \succeq \tval$, i.e., the evaluation of $\varphi$ with respect to $k$ on $\lambda(\rho)$ determines the winner of the play~$\rho$. Further, a (winning) strategy is a finite-state strategy if there exists a finite-state machine computing it in the usual sense (see~\cite[Chapter~2]{GraedelThomasWilke02} for details).

We are interested in solving $\rprompt$ games, i.e., in solving the following problem.

\begin{problem} 
Let $\game$ be an $\rprompt$ game and $v$ a vertex. Determine whether Player~$0$ has a winning strategy for $\game$ from $v$ and compute a finite-state winning strategy if so.
\end{problem}

Again, our solution to this problem relies on Lemma~\ref{lemma-prltl2pltl} and the fact that solving $\prompt$ games is in $\twoexp$~\cite{KupfermanPitermanVardi09,Zimmermann13}.

\begin{theorem} \label{thm-rpltlgames}
Solving $\rprompt$ games is $\twoexp$-complete. 	
\end{theorem}

Here we have a matching lower bound, as solving games with $\ltl$ conditions without next, until, and release is already $\twoexp$-hard~\cite{gameswithboxes}.

\section{Robust Linear Dynamic Logic}
\label{sec-rldl}
Next, we \myquot{robustify} $\ldl$ by generalizing the ideas underlying robust $\ltl$ to $\ldl$, obtaining the logic $\rldl$.  Again, following the precedent of robust $\ltl$, we equip robust operators with dots to distinguish them from non-robust ones.
The formulas of $\rldl$ are given by the grammar
\[
\varphi \cceq p \mid \neg \varphi \mid \varphi \wedge \varphi \mid \varphi \vee \varphi \mid \varphi \implies \varphi \mid  \ddiamonddot{r} \varphi 
  \mid \bboxdot{r} \varphi \qquad\qquad
    r  \cceq \phi \mid \varphi? \mid r+r \mid r \conc r \mid r^*,
\]
where $p$ ranges over the atomic propositions in $P$ and $\phi$ over propositional formulas over $P$. We refer to formulas of the form~$\ddiamonddot{r}\varphi$ and $\bboxdot{r}\varphi$ as diamond formulas and box formulas, respectively. In both cases, $r$ is the guard of the operator. An atom~$\varphi?$ of a regular expression is a test. We use the abbreviations~$\ttrue = p \vee \neg p$ and $\ffalse = p \wedge \neg p$ for some $p \in P$ and note that both are formulas and guards.
We denote the set of subformulas of $\varphi$ by $\cl(\varphi)$. Guards are not subformulas, but the formulas appearing in the tests are, e.g., we have $\cl(\ddiamonddot{\halfthinspace p?\conc q}p') = \set{\halfthinspace p, p', \ddiamonddot{\halfthinspace p?\conc q}p'}$. The size~$\card{\varphi}$ of $\varphi$ is the sum of $\card{\cl(\varphi)}$ and the sum of the lengths of the guards appearing in $\varphi$ (counted with multiplicity and measured in the number of operators).

Before we introduce the semantics of $\rldl$ we first recall the semantics of the robust always operator~$\Boxdot\varphi$ in robust $\ltl$. To this end, call a position~$j$ of a trace $\varphi$-satisfying if the suffix starting at position~$j$ satisfies $\varphi$. Now, the robust semantics are based on the following five cases, where the latter four distinguish various degrees of violating the formula~$\Box\varphi$: either all positions are $\varphi$-satisfying ($\Box$), almost all positions are $\varphi$-satisfying ($\Diamond\Box$), infinitely many positions are $\varphi$-satisfying ($\Box\Diamond$), some position is $\varphi$-satisfying ($\Diamond$), or no position is $\varphi$-satisfying.

A similar approach for a  formula~$\bboxdot{r}\varphi$ would be to consider the following possibilities, where a position~$j$ of a trace~$w$ is an $r$-match if the prefix of $w$ up to and including position $j-1$ is in the language of $r$: all $r$-matches are $\varphi$-satisfying, almost all $r$-matches are $\varphi$-satisfying, infinitely many $r$-matches are $\varphi$-satisfying, some $r$-match is $\varphi$-satisfying, or no $r$-match is $\varphi$-satisfying.
On a trace~$w$ with infinitely many $r$-matches, this is the natural generalization of the robust semantics. A trace, however, may only contain finitely many $r$-matches, or none at all. In the former case, there are not infinitely many $\varphi$-satisfying $r$-matches, but all $r$-matches could satisfy $\varphi$. Thus, the monotonicity of the cases is violated. 
We overcome this by interpreting \myquot{almost all} as \myquot{all} and \myquot{infinitely many} as \myquot{some} if there are only finitely many $r$-matches.\footnote{\label{footnote-altsemantics}There is an alternative definition inspired by the semantics of $\ltl$ on finite traces: Here, both $\Diamond\Box\varphi$ and $\Box\Diamond\varphi$ are equivalent to \myquot{$\varphi$ holds at the last position}. This suggests interpreting \myquot{almost all $r$-matches are $\varphi$-satisfying} and \myquot{infinitely many $r$-matches are $\varphi$-satisfying} as \myquot{the last $r$-match is $\varphi$-satisfying} in case there are only finitely many $r$-matches. Arguably, this definition is less intuitive than the one we propose to pursue.}

Also, the guard~$r$ may contain tests, which have to be evaluated to determine whether a position is an $r$-match. For this, we have to use the appropriate semantics for the robust box operator. For example, if we interpret $\bboxdot{r}\varphi$ to mean \myquot{almost all $r$-matches satisfy $\varphi$}, then the robust box operators in tests of $r$ are evaluated with this interpretation as well. This may, however, violate monotonicity (see Example~\ref{example-monotonicityviolation}), which we therefore hardcode in the semantics.

We now formalize the informal description above and subsequently show that this formalization satisfies all desired properties. To this end, we again define an evaluation function~$\rldleval$ mapping a trace~$w$ and a formula~$\varphi$ to a truth value. Also, we again denote the projection of $\rldleval(w,\varphi)$ to its $i$-th bit by $\rldleval_i(w,\varphi)$. For atomic propositions and Boolean connectives, the definition is the same as for $\rprompt$ introduced above (ignoring the bound~$k$) and for negation and implication, the definition is the same as for robust $\ltl$ (cf.~\cite{TabuadaNeider16}):

\begin{itemize}
\item $\rldleval(w,\neg \varphi) = \begin{cases} 
 	0000 &\text{if $\rldleval(w,\varphi) = 1111$,}\\ 
 	1111 &\text{if $\rldleval(w,\varphi) \neq 1111$,} 
\end{cases}$ and
 	 \item $\rldleval(w, \varphi_0 \implies \varphi_1) = \begin{cases}
 	1111 &\text{if $\rldleval(w,\varphi_0) \preceq \rldleval(w, \varphi_1)$,}\\
 	\rldleval(w, \varphi_1) &\text{if $\rldleval(w,\varphi_0) \succ \rldleval(w, \varphi_1)$.}
 \end{cases}$
\end{itemize}
To define the semantics of the diamond and the box operator, we need to first define the semantics of the guards: The match set~$\robRexp_i(w,r) \subseteq \nats$ for $i \in \set{1,2,3,4}$ contains all positions~$j$ of $w$ such that $\pref{w}{j}$ matches $r$ and is defined inductively as follows: 
\begin{itemize}

\item $\robRexp_i(w,\phi) = \set{1}$ if $w(0) \models \phi$ and $\robRexp_i(\phi,w) =  \emptyset$ otherwise, for propositional~$\phi$.

\item $\robRexp_i(w,\varphi?) = \set{0}$ if $\rldleval_i(w, \varphi)=1$ and $\robRexp_i(w,\varphi?) =  \emptyset$ otherwise.

\item $\robRexp_i(w,r_0 + r_1) = \robRexp_i(w,r_0) \cup \robRexp_i(w,r_1)$.

\item $\robRexp_i(w,r_0 \conc r_1) = \set{j_0 + j_1 \mid j_0, j_1 \ge 0 \text{ and } j_0 \in \robRexp_i(w,r_0) \text{ and } j_1 \in \robRexp_i(\suff{w}{j_0}, r_1)}$, i.e., for $j$ to be in $\robRexp_i(w,r_0 \conc r_1)$, it has to be the sum of natural numbers $j_0$ and $j_1$ such that $w$ has a prefix of length $j_0$ that matches $r_0$ and $\suff{w}{j_0}$ has a prefix of length~$j_1$ that matches~$r_1$. 

\item $\robRexp_i(w,r^*) = \set{0} \cup \set{j_1 + \cdots + j_\ell \mid 0 \le j_{\ell'} \in \robRexp_i(\suff{w}{j_1 + \cdots + j_{\ell'-1}},r) \text{ for all } \ell'\in\set{1, \ldots, \ell}}$, where we use $j_1 + \cdots + j_{0} = 0$. Thus, for $j$ to be in $\robRexp_i(w,r^*)$, it has to be expressible as $j = j_1 +\cdots + j_\ell$ with non-negative~$j_{\ell'}$ such that the prefix of $w$ of length~$j_1$ matches $r$, the prefix of length~$j_2$ of $\suff{w}{j_1}$ matches $r$, and in general, the prefix of length~$j_{\ell'}$ of $\suff{w}{j_1+\cdots +j_{\ell'-1}}$ matches $r$, for every $\ell' \in \set{1, \ldots, \ell}$.

\end{itemize}
Due to tests, membership of $j$ in $\robRexp_i(w,r)$ does, in general, not only depend on the prefix~$\pref{w}{j}$, but on the complete trace~$w$.
Also, the semantics of the propositional atom~$\phi$ differs from the semantics of the test~$\phi?$: the former consumes an input letter, while the latter one does not.
Thus, $\rldl$ (as $\ldl$) features both kinds of atoms. 
We define the intuition given above via  
\begin{itemize}
\item $\rldleval(w,  \ddiamonddot{r}\varphi) = b_1 b_2 b_3 b_4$ where 
$b_i = \max\nolimits_{j \in \robRexp_i(w,r)} \rldleval_i(\suff{w}{j}, \varphi)$ and
\item $\rldleval(w, \bboxdot{r}\varphi) = b_1 b_2 b_3 b_4$ with $b_i = \max\set{b_1', \ldots, b_i'}$ for every $i \in \set{ 1,2,3,4}$, where
\begin{itemize}
	
	\item $b_1' = \min_{j \in \robRexp_1(w,r)} \rldleval_1(\suff{w}{j},\varphi)$,
	
	\item $b_2' = \begin{cases}
	\max_{j' \in \nats} \min_{j \in \robRexp_2(w,r) \cap \set{j', j'+1, j'+2, \ldots}} \rldleval_2(\suff{w}{j},\varphi)&\text{if $\size{\robRexp_2(w,r)} = \infty$},\\
	
	\min_{j \in \robRexp_2(w,r)} \rldleval_2(\suff{w}{j},\varphi)&\text{if $0 < \size{\robRexp_2(w,r)} < \infty$},\\
	
	1 &\text{if $\size{\robRexp_2(w,r)} = 0$},
	\end{cases}$
	
	\item $b_3' = \begin{cases}
	\min_{j' \in \nats} \max_{j \in \robRexp_3(w,r) \cap \set{j', j'+1, j'+2, \ldots}} \rldleval_3(\suff{w}{j},\varphi)&\text{if $\size{\robRexp_3(w,r)} = \infty$},\\
	
	\max_{j \in \robRexp_3(w,r)} \rldleval_3(\suff{w}{j},\varphi)&\text{if $0 < \size{\robRexp_3(w,r)} < \infty$},\\
	
	1 &\text{if $\size{\robRexp_3(w,r)} = 0$},
	\end{cases}$
	
	\item $b_4' =\begin{cases} \max_{j \in \robRexp_4(w,r)} \rldleval_4(\suff{w}{j},\varphi) &\text{if $\size{\robRexp_4(w,r)} > 0$,}\\
1&\text{if $\size{\robRexp_4(w,r)} = 0$.}
\end{cases}
$
\end{itemize}
\end{itemize}

To give an intuitive description of the semantics, let us first generalize the notion of $r$-matches and $\varphi$-satisfiability.
We say that a position~$j$ of $w$ is an $r$-match of degree~$\tval$ if $j \in \robRexp_i(w,r)$ for the unique $i$ with $\tval = \itotruthvalue{i}$, which requires all tests in~$r$ to be evaluated w.r.t.\ $\rldleval_i$ (i.e., to some truth value at least $\tval$). Similarly, we say that a position~$j$ of $w$ is $\varphi$-satisfying of degree~$\tval$ if $\rldleval(\suff{w}{j} ,\varphi) \succeq \tval$, or if, equivalently, $\rldleval_i(\suff{w}{j},\varphi) =1$ for the unique $i$ with $\tval = \itotruthvalue{i}$.

Now, consider the $b_i'$ defining the semantics of the robust box operator: We have $b_1' = 1$ if all $r$-matches of degree~$1111$ are $\varphi$-satisfying of degree~$1111$. This is in particular satisfied if there is no such match.
Further, if there are infinitely (finitely) many $r$-matches of degree~$0111$, then $b_2' =1$ if almost all (if all) those matches are $\varphi$-satisfying of degree~$0111$. 
Dually, if there are infinitely (finitely) many $r$-matches of degree~$0011$, then $b_3' =1$ if infinitely many (at least one) of those matches are (is) $\varphi$-satisfying of degree~$0011$.
Finally, if there is at least one $r$-match of degree~$0001$, then $b_4' =1$ if at least one of those matches is $\varphi$-satisfying of degree~$0001$.
The cases where there is no $r$-match are irrelevant due to monotonicity, so we hardcode them to $1$.

\begin{example}
Consider the formula~$\bboxdot{r}q \implies \bboxdot{\ttrue\conc r}p$ with $r = (\ttrue; \ttrue)^*$, which expresses that the degree of violation of $q$ at \emph{even} positions should at most be the degree of violation of $p$ at \emph{odd} positions. Such a property cannot be expressed in $\rltl$, as even $\bboxdot{r}q$ is known to be inexpressible in $\ltl$~\cite{BaierKatoen08}.
\end{example}

First, we state that the semantics is well-defined. This is not obvious due to the case distinctions and the use of the matching sets~$\robRexp_i$ for different $i$.

\begin{lemma}
\label{lemma-semanticswelldefineds}
We have $\rldleval(w, \varphi) \in \bool_4$ for every trace~$w$ and every formula~$\varphi$.
\end{lemma}

To conclude the definition of the semantics, we give an example witnessing that the maximization over the $b_i'$ in the semantics of the box operator is indeed necessary to obtain monotonicity.

\begin{example}
\label{example-monotonicityviolation}
Let $\varphi = \bboxdot{r}\ffalse$ with $r = (\bboxdot{\ttrue^*}p)?$.
Moreover, consider the trace~$w = \emptyset\set{p}^\omega$. 
Then, we have $\rldleval(w, \bboxdot{\ttrue^*}p) = 0111$ and consequently $\robRexp_1(w,r) = \emptyset$ and $\robRexp_2(w,r) = \set{0}$. Therefore, $\min_{j \in \robRexp_1(w,r)} \rldleval_1(\suff{w}{j},\ffalse) = \min \emptyset = 1$, but $\min_{j \in \robRexp_2(w,r)} \rldleval_2(\suff{w}{j},\ffalse) = \min \set{0} = 0$. Thus, the bits~$b_1'$ and $b_2'$ inducing  $\rldleval(w, \bboxdot{r}\ffalse)$ are not monotonic, which explains the need to maximize  over the $b_i'$ to obtain the semantics of the robust box operator.
The traces~$(\emptyset \set{p})^\omega$ and~$\set{p} \emptyset^\omega$ witness that monotonicity can also be violated for the pairs $b_2',b_3'$ and $b_3',b_4'$.
\end{example}

We prove that $\rldl$ has the exponential compilation property. This allows us to solve the model checking and the synthesis problem using well-known and efficient automata-based algorithms. Furthermore, we are able to show that the complexity of these algorithms is asymptotically the same as the complexity of the algorithms for plain $\ldl$ and $\ltl$. 
In the terminology introduced in the introduction, we present a direct translation, i.e., we translate $\rldl$ directly into automata. 

\begin{theorem}
\label{theorem-translation-oldcor}
Let $\varphi$ be an $\rldl$ formula, $n = \size{\varphi}$, and $\tval \in \bool_4$.
There is a non-deterministic Büchi automaton~$\autb_{\varphi, \tval}$ with $2^{\bigo(n \log n)}$ states recognizing the language~$\set{w \in (\pow{P})^\omega \mid \rldleval(w,\varphi) \succeq \tval}$.

\end{theorem}

In order to obtain the desired Büchi automata, we follow the approach by Faymonville and Zimmermann~\cite{FaymonvilleZimmermann17}, who presented a bottom-up translation of parametric $\ldl$, an extension of $\ldl$ with prompt temporal operators, into alternating parity automata of linear size. These are then translated into Büchi automata of exponential size.  Here, we do not have to deal with prompt operators, but instead with the consequences of the five-valued semantics. Formally, we show that for every $\rldl$ formula~$\varphi$ and every $\tval \in \bool_4$, there is an alternating parity automaton~$\aut_{\varphi, \tval}$ with $\bigo(\size{\varphi})$ states recognizing the language~$\set{w \in (\pow{P})^\omega \mid \rldleval(w,\varphi) \succeq \tval} $.

As alternating parity automata are closed under union, intersection, and complementation, we directly obtain constructions for robust disjunction and conjunction, as these are defined with respect to the order of the truth values. Furthermore, even the robust semantics of implication and negation can be expressed using union, intersection, and complementation of automata. Thus, the only interesting cases are formulas of the form $\ddiamond{r}\varphi$ and $\bbox{r}\varphi$.  Faymonville and Zimmermann showed that one can translate~$r$ (which may contain tests) into an equivalent non-deterministic finite automaton with tests, i.e., states may be marked with formulas and the semantics of the automaton takes these into account. 

Fix some $\tval \in \bool_4$. One can take an automaton~$\aut_r$ for $r$, an alternating automaton~$\aut_{\varphi,\tval}$ for $\varphi$, and an alternating automaton~$\aut_{\theta,\tval}$ for each test~$\theta$ occurring in $r$, and combine them into an alternating automaton for $\ddiamond{r}\varphi$ with respect to $\tval$ that works as follows. It simulates $\aut_r$ and spawns a copy of $\aut_{\theta,\tval}$ each time a state marked by the test~$\theta$ is traversed. Furthermore, the acceptance condition is chosen such that the simulation has to be stopped at some accepting state of $\aut_r$, which implies that the prefix read so far is an $r$-match of degree~$\tval$. Additionally, when stopping the simulation of $\aut_r$, we additionally spawn a copy of $\aut_{\varphi,\tval}$ to check that this $r$-match is $\varphi$-satisfying of degree~$\tval$. Altogether, the resulting automaton checks that there is an $r$-match of degree~$\tval$ that is $\varphi$-satisfying of degree~$\tval$, i.e., it is indeed equivalent to $\ddiamond{r}\varphi$ with respect to~$\tval$.  

The construction for a formula of the form~$\bbox{r}\varphi$ relies on dual arguments, but is more involved due to the case distinctions in the definition of the robust semantics of the box operator. Using standard arguments about infinite languages of finite words, one can show that each of the conditions on $\robRexp_i$ used in the case distinctions can be checked by an automaton obtained from $\aut_r$. Furthermore, one can construct alternating automata checking that all (almost all, infinitely many, some) $r$-matches of some degree~$\tval$ are $\varphi$-satisfying of degree~$\tval$ by dualizing the construction for diamond formulas sketched above. Here, one heavily relies on alternation and the parity acceptance condition allowing to express finiteness and infiniteness properties. Finally, the case distinctions themselves can be implemented using the closure properties of alternating automata. We present the full construction in the full version~\cite{fullversion}.

Furthermore, as it is done for the similar construction for $\pldl$~\cite{FaymonvilleZimmermann17}, one can show that the automata can indeed be constructed efficiently: the non-deterministic Büchi automaton~$\autb_{\varphi, \tval}$  can be constructed on-the-fly in polynomial space, which is crucial to obtain a model checking algorithm with polynomial space requirements.

\subsection{Expressiveness}
\label{subsec-rldl-expressiveness}
In this section, we compare the expressiveness of $\rldl$ to that of $\rltl$ and $\ldl$. Following Tabuada and Neider~\cite{TabuadaNeider16} we focus on the fragment~$\rltl$ without next, until and release operators. While the next and until operator could be added easily, the robust semantics of the release operator is incompatible with our definition of the robust box operator. 
 It turns out as expected, that $\rldl$ subsumes $\rltl$. Conversely, every $\rldl$ formula~$\varphi$ can be translated into four $\ldl$ formulas~$\varphi_1, \ldots, \varphi_4$ that encode $\varphi$ in the following sense: We have $\rldleval_i(w,\varphi) = \ldleval(w, \varphi_i)$ for every $w$. 

\begin{theorem}
\label{thm-fragments}
Both $\rltl$ and $\ldl$ can be embedded into $\rldl$. 
\end{theorem}

As $\ltl$ is a syntactic fragment of $\ldl$, we immediately obtain that $\ltl$ can be embedded into $\rldl$ and, thus, $\rldl$ inherits the lower bounds of $\ltl$. 

Our next theorem states that $\ldl$ and $\rldl$ are of equal expressiveness. 
The direction from $\ldl$ to $\rldl$ was shown in Theorem~\ref{thm-fragments}, hence we focus on the other one.
Following Tabuada and Neider~\cite{TabuadaNeider16}, we construct for every $\rldl$ formula~$\varphi$ four $\ldl$ formulas~$\varphi_1, \ldots, \varphi_4$ encoding $\varphi$ as explained above. The construction relies on Theorem~\ref{theorem-translation-oldcor}, unlike the analogous result translating robust $\ltl$ directly into $\ltl$~\cite{TabuadaNeider16}.

\begin{theorem}\label{thm:rLDL-LDL-equally-expressive}
$\ldl$ and $\rldl$ are equally expressive and the translations are effective.
\end{theorem}

In general, translating an $\rldl$ formula into an equivalent $\ldl$ formula incurs a triply-exponential blow-up when using the translation described in the proof. On a more positive note, the resulting $\ldl$ formula is test-free, i.e., it does not contain tests in its guards.
We leave the question of whether there are non-trivial lower bounds on the translation for future work. For the special case of translating~$\rltl$ into $\ltl$ mentioned above, there is only a linear blowup. This translation was presented by Tabuada and Neider~\cite{TabuadaNeider16}, but they only claimed an exponential upper bound. However, closer inspection shows that it is linear if the size of formulas is measured in the number of distinct subformulas, not the length of the formula.

\subsection{Model Checking and Synthesis}
\label{subsec-rldl-modelchecking}

Theorem~\ref{thm:rLDL-LDL-equally-expressive} immediately provides solutions for typical applications of $\rldl$, such as model checking and synthesis, by reducing the problem from the domain of $\rldl$ to that of $\ldl$. However, the price to pay for this approach is a triply-exponential blow-up in the size of the resulting $\ldl$ formula, which is clearly prohibitive for any real-world application. For this reason, we now develop more efficient model checking and synthesis techniques that are based on our direct translation of $\rldl$ into automata (Theorem~\ref{theorem-translation-oldcor}).


We begin with the $\rldl$ model checking checking problem, which is defined as follows.

\begin{problem} \label{prob:rLDL-model-checking}
Let $\varphi$ be an $\rldl$ formula, $\sys$ a transition system, and let $\tval \in \bool_4$. Does $\rldleval(\trace(\rho), \varphi) \succeq \tval$ hold true for all paths~$\rho \in \Pi_\sys$?
\end{problem}

The exponential compilation property (see Theorem~\ref{theorem-translation-oldcor}) and standard on-the-fly techniques for checking emptiness of exponentially-sized Büchi automata~\cite{VardiWolper94} yield a $\pspace$ upper bound on the complexity of Problem~\ref{prob:rLDL-model-checking}. The matching lower bound follows from the subsumption of $\ldl$ shown above, as model checking $\ldl$ is $\pspace$-complete.

\begin{theorem} \label{theorem:rldl-model-checking-complexity}
$\rldl$ model checking is $\pspace$-complete.
\end{theorem}

Similar to model checking, the translation from $\rldl$ formulas to automata provides us with an effective means to synthesize reactive controllers from $\rldl$ specifications, i.e., for the following problem, where an $\rldl$ game now has the form~$(\ggraph, \varphi, \beta)$ and Player~$0$ wins a play if and only if its trace~$w$ satisfies $\rldleval(w, \varphi) \ge \beta$.

\begin{problem} \label{prob:rLDL-synthesis}
Let $\game$ be an $\rldl$ game and $v$ a vertex.  Determine whether Player~$0$ has a winning strategy for $\game$ from $v$ and compute a finite-state winning strategy if so.
\end{problem}

Theorem~\ref{theorem-translation-oldcor} provides a straightforward way to solve Problem~\ref{prob:rLDL-synthesis} by reducing it to solving classical parity games (again, see~\cite[Chapter~2]{GraedelThomasWilke02} for an introduction to parity games) while the lower bound follows from the subsumption of $\ldl$.

\begin{theorem}\label{theorem-rldlgames}
Solving $\rldl$ games is $\twoexp$-complete.
\end{theorem}

\section{Towards Robust and Prompt Linear Dynamic Logic}
\label{sec-towardsrpldl}
In the previous sections, we studied robust $\ldl$, i.e., we combined robustness and increased expressiveness, and robust $\prompt$, i.e., we combined robustness and quantitative operators. The third combination of two aspects, i.e., quantitative operators and increased expressiveness, has been studied before~\cite{FaymonvilleZimmermann17}. For all three resulting logics, model checking and synthesis have the same complexity as for plain $\ltl$. 

Here, we consider the combination of all three extensions, obtaining the logic $\rpromptldl$, robust $\promptldl$. The syntax is obtained by adding the prompt diamond operator~$\promptddiamonddot{r}\varphi$ to $\ldl$, by restricting negations to atomic formulas, and by disallowing implications. Here, $r$ is a guard as in $\rldl$, which may contain tests, i.e., formulas of $\rpromptldl$.
Similarly, the semantics is defined as expected, i.e., it is obtained by extending the semantics of $\rldl$ with a bound~$k$ for the prompt diamond operator~$\promptddiamonddot{r}\varphi$. Now, its semantics  requires the existence of a $\varphi$-satisfying $r$-match within the next $k$ steps. Formal definitions are as expected and presented in the full version~\cite{fullversion}.

\begin{example}
\label{example-rpromptldl}
Consider the formula~$\bboxdot{((\neg t)^*\conc t\conc (\neg t)^*\conc t )^*} \promptddiamonddot{\ttrue^*} s $ and interpret $t$ as the tick of a clock and $s$ as a synchronization. Then, the formula intuitively expresses that every other tick of the clock is followed after a bounded number of steps (not ticks!) by a synchronization. 

More formally, the different degrees of satisfaction of $\varphi$ express the following possibilities, with respect to a given bound~$k$:
(i) every even clock tick is followed by a synchronization within $k$ steps;
(ii) almost every even clock tick is followed by a synchronization within $k$ steps;
(iii) infinitely many even clock ticks are followed by a synchronization within $k$ steps;
(iv) there is at least one even clock tick that is followed by a synchronization within $k$ steps.

This property can neither be expressed in (robust) $\ldl$ nor in (robust) $\prompt$. Also note that unlike for the similar formula from Example~\ref{example-rprompt}, the last two possibilities are not trivial, as we now only consider positions with an even clock tick and not all positions. 
\end{example}

In the previous sections, we have seen two approaches to translating robust logics into Büchi automata, the direct and the reduction-based one. Both are extensions of translations originally introduced by Tabuada and Neider for robust $\ltl$. The former one translates a formula of a robust logic directly into an equivalent Büchi automaton while the latter one first translates a formula of a robust logic  into an equivalent classical (non-robust) logic, for which a translation into equivalent Büchi automata is already known. For robust $\ltl$, both approaches are applicable~\cite{TabuadaNeider16} and yield Büchi automata of exponential size. 
Here, out of necessity, we apply both approaches: for robust $\ldl$, we present a direct translation while we present a reduction-based approach for robust $\prompt$. Let us quickly elaborate the reasons for this.

First, consider the reduction-based approach for robust $\ltl$, which translates a formula~$\varphi$ of robust $\ltl$ and a truth value~$\beta \succ 0000$ into an $\ltl$ formula~$\varphi_\beta$ that captures $\varphi$ with respect to $\beta$. To this end, the formula~$\varphi_\beta$ implements the intuitive meaning of the robust semantics for the always operator, e.g., we have $(\Boxdot p )_{1111} = \Box p$, $(\Boxdot p )_{0111} = \Diamond\Box p$, $(\Boxdot p )_{0011} = \Box\Diamond p$, and $(\Boxdot p )_{0001} = \Diamond p$. 

Trying to apply this approach to the  $\rldl$ formula~$\varphi = \bboxdot{r} p$, say for $\beta = 0111$, would imply using a formula of the form~$\ddiamonddot{r_0} \bboxdot{r_1} p$ where $r_0$ and $r_1$ are obtained by \myquot{splitting} up $r$. It captures the robust semantics of $\varphi$ with respect to $\beta$ on some trace~$w$ by expressing that there is an $r_0$-match~$j$ such that every $r_1$-match in $\suff{w}{j}$ is $p$-satisfying with degree~$\beta$. Thus, $r_0$ and $r_1$ have to be picked such that the $r_1$-matches in $\suff{w}{j}$ as above correspond exactly to the $r$-matches in $w$.
Further, to obtain a translation of optimal complexity, $r_0$ and $r_1$ have to be of polynomial size in $\size{r}$. It is an open problem whether such a splitting is always possible, in particular in the presence of tests in $r$ and guards with only finitely many $r$-matches.

Secondly, recall that the direct approach to robust $\ltl$ translates a formula~$\varphi$ of $\rltl$ into a Büchi automaton that captures $\varphi$ with respect to all $\beta \in \bool_4$ (by considering five initial states, one for each $\beta$). Trying to apply this approach to robust $\prompt$ requires using a more general automaton model that is able to capture the quantitative nature of the prompt diamond operator while still yielding a model checking and a synthesis algorithm with the desired complexity. To the best of our knowledge, no such translation from $\prompt$ to automata has been presented in the literature, which would be a special case of our construction here.

Thus, according to the state-of-the-art, the direct approach is the only viable one for robust extensions of $\ldl$ while the reduction-based approach is the only viable one for robust extensions of $\prompt$. This leaves us with no viable approach for $\rpromptldl$. 

Nevertheless, in the full version~\cite{fullversion}, we present a fragment of $\rpromptldl$ and a reduction-based translation to $\promptldl$ for it. The fragment is obtained by disallowing tests in guards and requiring them to always have infinitely many matches. For such formulas, one can translate the guard into a deterministic finite automaton (without tests) and then use this automaton to \myquot{split} $r$. However, this involves multiple exponential blowups and hence does not prove that the fragment has the exponential compilation property. Nonetheless, this translation shows that both model checking and synthesis are decidable for this fragment. The decidability of these problems for full $\rpromptldl$ is left for further research and seemingly requires new approaches.

\section{Conclusion}
\label{sec-conc}
We addressed the problems of verification and synthesis with robust, expressive, and quantitative linear temporal specifications.
Inspired by robust $\ltl$, we have first developed robust extensions of the logics $\ldl$ and $\prompt$, named $\rldl$ and $\rprompt$, respectively.
Then, we combined $\rldl$ and $\rprompt$ into a third logic, named $\rpromptldl$, which has the expressiveness of $\omega$-regular languages and allows robust reasoning about timing bounds.

For~$\rldl$ and~$\rprompt$, we have shown how to solve the model checking and synthesis problem relying on the exponential compilation property. Hence, all these problems are not harder than those for plain $\ltl$. 
The situation for the combination of all three basic logics, i.e., for $\rpromptldl$, is less encouraging. In the full version~\cite{fullversion}, we show the problems to be decidable for an important fragment, but due to a blowup of the formulas during the reduction, we (most likely) do not obtain optimal algorithms. Decidability for the full logic remains open. 

In future work, we aim to determine the exact complexity of the model checking and synthesis problem for (full) $\rpromptldl$. One promising approach is to generalize the translation of $\rldl$ into alternating parity automata. However, this requires a suitable quantitative alternating automata model with strong closure properties that can be transformed into equivalent non-deterministic and deterministic automata. 

Another promising direction for further research is to study the semantics for the robust box operator proposed in Footnote~\ref{footnote-altsemantics} on Page~\pageref{footnote-altsemantics}. In particular, it is open whether the translation into alternating automata can be generalized to this setting without a blowup.
Also, we leave open whether full robust $\ltl$, i.e., with until and release, can be embedded into $\rldl$. As is, the robust semantics of the release operator (see~\cite{TabuadaNeider16}) is not compatible with our robust semantics for $\rldl$. In future work, we plan to study generalizations of full robust $\ltl$. 

Another natural question is whether the techniques developed for $\rldl$ can be applied to a robust version of the Property Specification Language~\cite{EisnerFismanPSL}. 

\bibliographystyle{eptcs}
\bibliography{biblio}

\newcommand{\noopsort}[1]{}
\begin{thebibliography}{10}
\providecommand{\bibitemdeclare}[2]{}
\providecommand{\surnamestart}{}
\providecommand{\surnameend}{}
\providecommand{\urlprefix}{Available at }
\providecommand{\url}[1]{\texttt{#1}}
\providecommand{\href}[2]{\texttt{#2}}
\providecommand{\urlalt}[2]{\href{#1}{#2}}
\providecommand{\doi}[1]{doi:\urlalt{http://dx.doi.org/#1}{#1}}
\providecommand{\bibinfo}[2]{#2}

\bibitemdeclare{article}{AlurEtessamiLaTorrePeled01}
\bibitem{AlurEtessamiLaTorrePeled01}
\bibinfo{author}{Rajeev \surnamestart Alur\surnameend}, \bibinfo{author}{Kousha
  \surnamestart Etessami\surnameend}, \bibinfo{author}{Salvatore~La
  \surnamestart Torre\surnameend} \& \bibinfo{author}{Doron \surnamestart
  Peled\surnameend} (\bibinfo{year}{2001}): \emph{\bibinfo{title}{Parametric
  Temporal Logic for ``Model Measuring''}}.
\newblock {\sl \bibinfo{journal}{ACM Trans. Comput. Log.}}
  \bibinfo{volume}{2}(\bibinfo{number}{3}), pp. \bibinfo{pages}{388--407},
  \doi{10.1145/377978.377990}.

\bibitemdeclare{inproceedings}{gameswithboxes}
\bibitem{gameswithboxes}
\bibinfo{author}{Rajeev \surnamestart Alur\surnameend},
  \bibinfo{author}{Salvatore \surnamestart {La Torre}\surnameend} \&
  \bibinfo{author}{P.~\surnamestart Madhusudan\surnameend}
  (\bibinfo{year}{2003}): \emph{\bibinfo{title}{Playing Games with Boxes and
  Diamonds}}.
\newblock In \bibinfo{editor}{Roberto~M. \surnamestart Amadio\surnameend} \&
  \bibinfo{editor}{Denis \surnamestart Lugiez\surnameend}, editors: {\sl
  \bibinfo{booktitle}{{CONCUR} 2003}}, {\sl \bibinfo{series}{LNCS}}
  \bibinfo{volume}{2761}, \bibinfo{publisher}{Springer}, pp.
  \bibinfo{pages}{127--141}, \doi{10.1007/978-3-540-45187-7\_8}.

\bibitemdeclare{article}{AlurLatorre04}
\bibitem{AlurLatorre04}
\bibinfo{author}{Rajeev \surnamestart Alur\surnameend} \&
  \bibinfo{author}{Salvatore~La \surnamestart Torre\surnameend}
  (\bibinfo{year}{2004}): \emph{\bibinfo{title}{Deterministic generators and
  games for {LTL} fragments}}.
\newblock {\sl \bibinfo{journal}{ACM Trans. Comput. Log.}}
  \bibinfo{volume}{5}(\bibinfo{number}{1}), pp. \bibinfo{pages}{1--25},
  \doi{10.1145/963927.963928}.

\bibitemdeclare{book}{BaierKatoen08}
\bibitem{BaierKatoen08}
\bibinfo{author}{Christel \surnamestart Baier\surnameend} \&
  \bibinfo{author}{Joost-Pieter \surnamestart Katoen\surnameend}
  (\bibinfo{year}{2008}): \emph{\bibinfo{title}{Principles of Model Checking}}.
\newblock \bibinfo{publisher}{The MIT Press}.

\bibitemdeclare{inproceedings}{DBLP:conf/cav/BloemCGHJ10}
\bibitem{DBLP:conf/cav/BloemCGHJ10}
\bibinfo{author}{Roderick \surnamestart Bloem\surnameend},
  \bibinfo{author}{Krishnendu \surnamestart Chatterjee\surnameend},
  \bibinfo{author}{Karin \surnamestart Greimel\surnameend},
  \bibinfo{author}{Thomas~A. \surnamestart Henzinger\surnameend} \&
  \bibinfo{author}{Barbara \surnamestart Jobstmann\surnameend}
  (\bibinfo{year}{2010}): \emph{\bibinfo{title}{Robustness in the Presence of
  Liveness}}.
\newblock In: {\sl \bibinfo{booktitle}{{CAV} 2010}}, {\sl
  \bibinfo{series}{LNCS}} \bibinfo{volume}{6174},
  \bibinfo{publisher}{Springer}, pp. \bibinfo{pages}{410--424},
  \doi{10.1007/978-3-642-14295-6\_36}.

\bibitemdeclare{inproceedings}{DallalNeiderTabuada16}
\bibitem{DallalNeiderTabuada16}
\bibinfo{author}{Eric \surnamestart Dallal\surnameend}, \bibinfo{author}{Daniel
  \surnamestart Neider\surnameend} \& \bibinfo{author}{Paulo \surnamestart
  Tabuada\surnameend} (\bibinfo{year}{2016}): \emph{\bibinfo{title}{Synthesis
  of safety controllers robust to unmodeled intermittent disturbances}}.
\newblock In: {\sl \bibinfo{booktitle}{CDC 2016}}, pp.
  \bibinfo{pages}{7425--7430}, \doi{10.1109/CDC.2016.7799416}.

\bibitemdeclare{inproceedings}{GiacomoVardi13}
\bibitem{GiacomoVardi13}
\bibinfo{author}{Giuseppe \surnamestart {De Giacomo}\surnameend} \&
  \bibinfo{author}{Moshe~Y. \surnamestart Vardi\surnameend}
  (\bibinfo{year}{2013}): \emph{\bibinfo{title}{Linear Temporal Logic and
  Linear Dynamic Logic on Finite Traces}}.
\newblock In \bibinfo{editor}{Francesca \surnamestart Rossi\surnameend},
  editor: {\sl \bibinfo{booktitle}{IJCAI}}, \bibinfo{publisher}{IJCAI/AAAI}.
\newblock
  \urlprefix\url{http://www.aaai.org/ocs/index.php/IJCAI/IJCAI13/paper/view/6997}.

\bibitemdeclare{inproceedings}{DBLP:conf/formats/DonzeM10}
\bibitem{DBLP:conf/formats/DonzeM10}
\bibinfo{author}{Alexandre \surnamestart Donz{\'{e}}\surnameend} \&
  \bibinfo{author}{Oded \surnamestart Maler\surnameend} (\bibinfo{year}{2010}):
  \emph{\bibinfo{title}{Robust Satisfaction of Temporal Logic over Real-Valued
  Signals}}.
\newblock In \bibinfo{editor}{Krishnendu \surnamestart Chatterjee\surnameend}
  \& \bibinfo{editor}{Thomas~A. \surnamestart Henzinger\surnameend}, editors:
  {\sl \bibinfo{booktitle}{{FORMATS} 2010}}, {\sl \bibinfo{series}{LNCS}}
  \bibinfo{volume}{6246}, \bibinfo{publisher}{Springer}, pp.
  \bibinfo{pages}{92--106}, \doi{10.1007/978-3-642-15297-9\_9}.

\bibitemdeclare{inproceedings}{DBLP:conf/acsd/DoyenHLN10}
\bibitem{DBLP:conf/acsd/DoyenHLN10}
\bibinfo{author}{Laurent \surnamestart Doyen\surnameend},
  \bibinfo{author}{Thomas~A. \surnamestart Henzinger\surnameend},
  \bibinfo{author}{Axel \surnamestart Legay\surnameend} \&
  \bibinfo{author}{Dejan \surnamestart Nickovic\surnameend}
  (\bibinfo{year}{2010}): \emph{\bibinfo{title}{Robustness of Sequential
  Circuits}}.
\newblock In \bibinfo{editor}{Lu{\'{\i}}s \surnamestart Gomes\surnameend},
  \bibinfo{editor}{Victor \surnamestart Khomenko\surnameend} \&
  \bibinfo{editor}{Jo{\~{a}}o~M. \surnamestart Fernandes\surnameend}, editors:
  {\sl \bibinfo{booktitle}{{ACSD} 2010}}, \bibinfo{publisher}{{IEEE} Computer
  Society}, pp. \bibinfo{pages}{77--84}, \doi{10.1109/ACSD.2010.26}.

\bibitemdeclare{book}{EisnerFismanPSL}
\bibitem{EisnerFismanPSL}
\bibinfo{author}{C.~\surnamestart Eisner\surnameend} \&
  \bibinfo{author}{D.~\surnamestart Fisman\surnameend} (\bibinfo{year}{2006}):
  \emph{\bibinfo{title}{A Practical Introduction to PSL}}.
\newblock \bibinfo{series}{Integrated Circuits and Systems},
  \bibinfo{publisher}{Springer}, \doi{10.1007/978-0-387-36123-9}.

\bibitemdeclare{article}{DBLP:journals/tcs/FainekosP09}
\bibitem{DBLP:journals/tcs/FainekosP09}
\bibinfo{author}{Georgios~E. \surnamestart Fainekos\surnameend} \&
  \bibinfo{author}{George~J. \surnamestart Pappas\surnameend}
  (\bibinfo{year}{2009}): \emph{\bibinfo{title}{Robustness of temporal logic
  specifications for continuous-time signals}}.
\newblock {\sl \bibinfo{journal}{Theor. Comput. Sci.}}
  \bibinfo{volume}{410}(\bibinfo{number}{42}), pp. \bibinfo{pages}{4262--4291},
  \doi{10.1016/j.tcs.2009.06.021}.

\bibitemdeclare{article}{FaymonvilleZimmermann17}
\bibitem{FaymonvilleZimmermann17}
\bibinfo{author}{Peter \surnamestart Faymonville\surnameend} \&
  \bibinfo{author}{Martin \surnamestart Zimmermann\surnameend}
  (\bibinfo{year}{2017}): \emph{\bibinfo{title}{Parametric Linear Dynamic
  Logic}}.
\newblock {\sl \bibinfo{journal}{Inf. Comput.}} \bibinfo{volume}{253}, pp.
  \bibinfo{pages}{237--256}, \doi{10.1016/j.ic.2016.07.009}.

\bibitemdeclare{inproceedings}{Fix08}
\bibitem{Fix08}
\bibinfo{author}{Limor \surnamestart Fix\surnameend} (\bibinfo{year}{2008}):
  \emph{\bibinfo{title}{Fifteen Years of Formal Property Verification in
  Intel}}.
\newblock In \bibinfo{editor}{Orna \surnamestart Grumberg\surnameend} \&
  \bibinfo{editor}{Helmut \surnamestart Veith\surnameend}, editors: {\sl
  \bibinfo{booktitle}{25 Years of Model Checking - History, Achievements,
  Perspectives}}, {\sl \bibinfo{series}{LNCS}} \bibinfo{volume}{5000},
  \bibinfo{publisher}{Springer}, pp. \bibinfo{pages}{139--144},
  \doi{10.1007/978-3-540-69850-0\_8}.

\bibitemdeclare{proceedings}{GraedelThomasWilke02}
\bibitem{GraedelThomasWilke02}
\bibinfo{editor}{Erich \surnamestart Gr{\"a}del\surnameend},
  \bibinfo{editor}{Wolfgang \surnamestart Thomas\surnameend} \&
  \bibinfo{editor}{Thomas \surnamestart Wilke\surnameend}, editors
  (\bibinfo{year}{2002}): \emph{\bibinfo{title}{Automata, Logics, and Infinite
  Games: A Guide to Current Research}}. {\sl \bibinfo{series}{LNCS}}
  \bibinfo{volume}{2500}, \bibinfo{publisher}{Springer},
  \doi{10.1007/3-540-36387-4}.

\bibitemdeclare{article}{Koymans90}
\bibitem{Koymans90}
\bibinfo{author}{Ron \surnamestart Koymans\surnameend} (\bibinfo{year}{1990}):
  \emph{\bibinfo{title}{Specifying real-time properties with metric temporal
  logic}}.
\newblock {\sl \bibinfo{journal}{Real-Time Systems}} \bibinfo{volume}{2}, pp.
  \bibinfo{pages}{255--299}, \doi{10.1007/BF01995674}.

\bibitemdeclare{article}{KupfermanPitermanVardi09}
\bibitem{KupfermanPitermanVardi09}
\bibinfo{author}{Orna \surnamestart Kupferman\surnameend}, \bibinfo{author}{Nir
  \surnamestart Piterman\surnameend} \& \bibinfo{author}{Moshe~Y. \surnamestart
  Vardi\surnameend} (\bibinfo{year}{2009}): \emph{\bibinfo{title}{From Liveness
  to Promptness}}.
\newblock {\sl \bibinfo{journal}{Formal Methods in System Design}}
  \bibinfo{volume}{34}(\bibinfo{number}{2}), pp. \bibinfo{pages}{83--103},
  \doi{10.1007/s10703-009-0067-z}.

\bibitemdeclare{inproceedings}{LeuckerSanchez07}
\bibitem{LeuckerSanchez07}
\bibinfo{author}{Martin \surnamestart Leucker\surnameend} \&
  \bibinfo{author}{C{\'{e}}sar \surnamestart S{\'{a}}nchez\surnameend}
  (\bibinfo{year}{2007}): \emph{\bibinfo{title}{Regular Linear Temporal
  Logic}}.
\newblock In \bibinfo{editor}{Cliff~B. \surnamestart Jones\surnameend},
  \bibinfo{editor}{Zhiming \surnamestart Liu\surnameend} \&
  \bibinfo{editor}{Jim \surnamestart Woodcock\surnameend}, editors: {\sl
  \bibinfo{booktitle}{ICTAC 2007}}, {\sl \bibinfo{series}{LNCS}}
  \bibinfo{volume}{4711}, \bibinfo{publisher}{Springer}, pp.
  \bibinfo{pages}{291--305}, \doi{10.1007/978-3-540-75292-9\_20}.

\bibitemdeclare{inproceedings}{DBLP:conf/rtss/MajumdarS09}
\bibitem{DBLP:conf/rtss/MajumdarS09}
\bibinfo{author}{Rupak \surnamestart Majumdar\surnameend} \&
  \bibinfo{author}{Indranil \surnamestart Saha\surnameend}
  (\bibinfo{year}{2009}): \emph{\bibinfo{title}{Symbolic Robustness Analysis}}.
\newblock In \bibinfo{editor}{Theodore~P. \surnamestart Baker\surnameend},
  editor: {\sl \bibinfo{booktitle}{{RTSS} 2009}}, \bibinfo{publisher}{{IEEE}
  Computer Society}, pp. \bibinfo{pages}{355--363}, \doi{10.1109/RTSS.2009.17}.

\bibitemdeclare{article}{fullversion}
\bibitem{fullversion}
\bibinfo{author}{Daniel \surnamestart Neider\surnameend},
  \bibinfo{author}{Alexander \surnamestart Weinert\surnameend} \&
  \bibinfo{author}{Martin \surnamestart Zimmermann\surnameend}
  (\bibinfo{year}{2018}): \emph{\bibinfo{title}{Robust, Expressive, and
  Quantitative Linear Temporal Logics: Pick any Two for Free (full version)}}.
\newblock {\sl \bibinfo{journal}{arXiv}} \bibinfo{volume}{1808.09028}.
\newblock \urlprefix\url{http://arxiv.org/abs/1808.09028}.

\bibitemdeclare{inproceedings}{NeiderWeinertZimmermann18}
\bibitem{NeiderWeinertZimmermann18}
\bibinfo{author}{Daniel \surnamestart Neider\surnameend},
  \bibinfo{author}{Alexander \surnamestart Weinert\surnameend} \&
  \bibinfo{author}{Martin \surnamestart Zimmermann\surnameend}
  (\bibinfo{year}{2018}): \emph{\bibinfo{title}{Synthesizing Optimally
  Resilient Controllers}}.
\newblock In \bibinfo{editor}{Dan~R. \surnamestart Ghica\surnameend} \&
  \bibinfo{editor}{Achim \surnamestart Jung\surnameend}, editors: {\sl
  \bibinfo{booktitle}{CSL 2018}}, {\sl \bibinfo{series}{LIPIcs}}
  \bibinfo{volume}{119}, \bibinfo{publisher}{Schloss Dagstuhl - LZI}, pp.
  \bibinfo{pages}{34:1--34:17}, \doi{10.4230/LIPIcs.CSL.2018.34}.

\bibitemdeclare{inproceedings}{Pnueli77}
\bibitem{Pnueli77}
\bibinfo{author}{Amir \surnamestart Pnueli\surnameend} (\bibinfo{year}{1977}):
  \emph{\bibinfo{title}{The temporal logic of programs}}.
\newblock In: {\sl \bibinfo{booktitle}{FOCS 1977}}, \bibinfo{publisher}{IEEE},
  pp. \bibinfo{pages}{46--57}, \doi{10.1109/SFCS.1977.32}.

\bibitemdeclare{article}{DBLP:journals/tac/TabuadaCRM14}
\bibitem{DBLP:journals/tac/TabuadaCRM14}
\bibinfo{author}{Paulo \surnamestart Tabuada\surnameend},
  \bibinfo{author}{Sina~Yamac \surnamestart Caliskan\surnameend},
  \bibinfo{author}{Matthias \surnamestart Rungger\surnameend} \&
  \bibinfo{author}{Rupak \surnamestart Majumdar\surnameend}
  (\bibinfo{year}{2014}): \emph{\bibinfo{title}{Towards Robustness for
  Cyber-Physical Systems}}.
\newblock {\sl \bibinfo{journal}{{IEEE} Trans. Automat. Contr.}}
  \bibinfo{volume}{59}(\bibinfo{number}{12}), pp. \bibinfo{pages}{3151--3163},
  \doi{10.1109/TAC.2014.2351632}.

\bibitemdeclare{inproceedings}{TabuadaNeider16}
\bibitem{TabuadaNeider16}
\bibinfo{author}{Paulo \surnamestart Tabuada\surnameend} \&
  \bibinfo{author}{Daniel \surnamestart Neider\surnameend}
  (\bibinfo{year}{2016}): \emph{\bibinfo{title}{Robust Linear Temporal Logic}}.
\newblock In \bibinfo{editor}{Jean{-}Marc \surnamestart Talbot\surnameend} \&
  \bibinfo{editor}{Laurent \surnamestart Regnier\surnameend}, editors: {\sl
  \bibinfo{booktitle}{{CSL} 2016}}, {\sl
  \bibinfo{series}{LIPIcs}}~\bibinfo{volume}{62}, \bibinfo{publisher}{Schloss
  Dagstuhl - LZI}, pp. \bibinfo{pages}{10:1--10:21},
  \doi{10.4230/LIPIcs.CSL.2016.10}.

\bibitemdeclare{inproceedings}{Vardi11}
\bibitem{Vardi11}
\bibinfo{author}{Moshe~Y. \surnamestart Vardi\surnameend}
  (\bibinfo{year}{2011}): \emph{\bibinfo{title}{The rise and fall of {LTL}}}.
\newblock In \bibinfo{editor}{Giovanna \surnamestart D'Agostino\surnameend} \&
  \bibinfo{editor}{Salvatore~La \surnamestart Torre\surnameend}, editors: {\sl
  \bibinfo{booktitle}{GandALF 2011}}, {\sl
  \bibinfo{series}{{EPTCS}}}~\bibinfo{volume}{54}.

\bibitemdeclare{article}{VardiWolper94}
\bibitem{VardiWolper94}
\bibinfo{author}{Moshe~Y. \surnamestart Vardi\surnameend} \&
  \bibinfo{author}{Pierre \surnamestart Wolper\surnameend}
  (\bibinfo{year}{1994}): \emph{\bibinfo{title}{Reasoning About Infinite
  Computations}}.
\newblock {\sl \bibinfo{journal}{Inf. Comput.}}
  \bibinfo{volume}{115}(\bibinfo{number}{1}), pp. \bibinfo{pages}{1--37},
  \doi{10.1006/inco.1994.1092}.

\bibitemdeclare{article}{Wolper83}
\bibitem{Wolper83}
\bibinfo{author}{Pierre \surnamestart Wolper\surnameend}
  (\bibinfo{year}{1983}): \emph{\bibinfo{title}{Temporal Logic Can Be More
  Expressive}}.
\newblock {\sl \bibinfo{journal}{Information and Control}}
  \bibinfo{volume}{56}(\bibinfo{number}{1/2}), pp. \bibinfo{pages}{72--99},
  \doi{10.1016/S0019-9958(83)80051-5}.

\bibitemdeclare{article}{Zimmermann13}
\bibitem{Zimmermann13}
\bibinfo{author}{Martin \surnamestart Zimmermann\surnameend}
  (\bibinfo{year}{2013}): \emph{\bibinfo{title}{Optimal Bounds in Parametric
  {LTL} Games}}.
\newblock {\sl \bibinfo{journal}{Theor. Comput. Sci.}} \bibinfo{volume}{493},
  pp. \bibinfo{pages}{30--45}, \doi{10.1016/j.tcs.2012.07.039}.

\bibitemdeclare{article}{Zimmermann18}
\bibitem{Zimmermann18}
\bibinfo{author}{Martin \surnamestart Zimmermann\surnameend}
  (\bibinfo{year}{2018}): \emph{\bibinfo{title}{Parameterized linear temporal
  logics meet costs: still not costlier than {LTL}}}.
\newblock {\sl \bibinfo{journal}{Acta Inf.}}
  \bibinfo{volume}{55}(\bibinfo{number}{2}), pp. \bibinfo{pages}{129--152},
  \doi{10.1007/s00236-016-0279-9}.

\end{thebibliography}
\end{document}